\documentclass[conference]{IEEEtran}
\IEEEoverridecommandlockouts

\usepackage{amsmath,amsfonts}
\usepackage{bm}
\usepackage{amssymb}
\usepackage{hhline}
\usepackage{algorithmic}
\usepackage{algorithm}
\usepackage{array}
\usepackage[dvipsnames]{xcolor}
\usepackage{multicol}
\usepackage{multirow}
\ifCLASSOPTIONcompsoc
    \usepackage[caption=false, font=normalsize, labelfont=sf, textfont=sf]{subfig}
\else
\usepackage[caption=false, font=footnotesize]{subfig}
\fi
\usepackage{textcomp}
\usepackage{stfloats}
\usepackage{hyperref}
\usepackage{verbatim}
\usepackage{graphicx}
\usepackage{cite}

\newcolumntype{P}[1]{>{\centering\arraybackslash}p{#1}}

\def\BibTeX{{\rm B\kern-.05em{\sc i\kern-.025em b}\kern-.08em
    T\kern-.1667em\lower.7ex\hbox{E}\kern-.125emX}}

\begin{document}

\title{A FUNQUE Approach to the Quality Assessment of Compressed HDR Videos \\
\thanks{This research was sponsored by a grant from Meta Video Infrastructure, and by grant number 2019844 for the National Science Foundation AI Institute for Foundations of Machine Learning (IFML).}
}

\author{\IEEEauthorblockN{Abhinau K. Venkataramanan\IEEEauthorrefmark{1}, Cosmin Stejerean\IEEEauthorrefmark{2}, Ioannis Katsavounidis\IEEEauthorrefmark{2}, Alan C. Bovik\IEEEauthorrefmark{1}}
\IEEEauthorblockA{\IEEEauthorrefmark{1}\textit{The University of Texas at Austin, Austin, USA} \\
\IEEEauthorrefmark{2}\textit{Meta Platforms, Inc., Menlo Park, USA} \\
abhinaukumar@utexas.edu, \{cstejerean, ikatsavounidis\}@meta.com, bovik@ece.utexas.edu}}

\maketitle

\begin{abstract}
Recent years have seen steady growth in the popularity and availability of High Dynamic Range (HDR) content, particularly videos, streamed over the internet. As a result, assessing the subjective quality of HDR videos, which are generally subjected to compression, is of increasing importance. In particular, we target the task of full-reference quality assessment of compressed HDR videos. The state-of-the-art (SOTA) approach HDRMAX involves augmenting off-the-shelf video quality models, such as VMAF, with features computed on non-linearly transformed video frames. However, HDRMAX increases the computational complexity of models like VMAF. Here, we show that an efficient class of video quality prediction models named FUNQUE+ achieves SOTA accuracy. This shows that the FUNQUE+ models are flexible alternatives to VMAF that achieve higher HDR video quality prediction accuracy at lower computational cost.
\end{abstract}

\begin{IEEEkeywords}
High Dynamic Range, Video Compression, FUNQUE, Perceptual Sensitivity
\end{IEEEkeywords}

\section{Introduction}
\label{sec:introduction}
The real world presents the human visual system with a wide range of luminances, i.e., brightness ranges, even in everyday settings. For example, the luminance of starlight is a mere 0.0003~cd/\(\text{m}^2\) (nits), while the luminance of bright sunlight on a clear day can reach 30,000 nits. Because of adaptive gain control mechanisms, particularly the iris' control of the pupil size, the visual system is able to perceive a wide range of brightnesses, from around \(10^{-6}\) nits to \(10^8\) nits. 

The iris control mechanism dominates over biochemical adaptation mechanisms when viewing videos since it can provide adaptivity within milliseconds, which the biochemical mechanisms cannot. The normal pupil size in adults varies from 2 to 4 mm in diameter in bright light to 4 to 8 mm in the dark. Since the amount of light that goes through the pupil is proportional to its surface, the iris can offer a factor of 16 gain control over the amount of light that reaches the retina. This factor, compounded with the native sensitivity of the cones in the fovea of our eyes, results in the high dynamic range that the human visual system enjoys when watching videos.

Until recently, widely deployed legacy imaging technologies could only capture the standard dynamic range (SDR) of brightnesses of about 100 nits. Likewise, SDR standards such as Rec.709 \cite{ref:rec_709} and sRGB \cite{ref:srgb} can only represent limited volumes of colors occupying only about a third of all visible colors \cite{ref:coverage}. By comparison, modern High Dynamic Range (HDR) standards such as BT.2100 PQ \cite{ref:pq} are capable of representing wide ranges of brightness, up to 10,000 nits. The PQ HDR standard is equipped with a wide color gamut that can represent nearly two-thirds of all visible colors \cite{ref:coverage}. Another HDR standard, called Hybrid-Log Gamma (HLG) \cite{ref:hlg}, supports a limited range of brightness (nominally, 1000 nits) \cite{ref:rec_2100} and was primarily designed for backward compatibility with SDR standards.

Due to the use of 10-bit and 12-bit representations, HDR videos require increased compression to achieve bitrate budgets stipulated by streaming internet bandwidth conditions, as compared to SDR video. Because of this, the development of objective video quality models that can be used to measure, monitor, and control the perceptual quality of compressed HDR videos has become important. 

Recent work on HDR quality modeling builds on a rich body of work targeting SDR video quality prediction. For example, the PU21 \cite{ref:pu21} non-linear transform was proposed as a perceptually uniform \cite{ref:lum2int} domain within which full-reference (FR) SDR video quality models may be adapted to HDR images and, by extension, videos. By using the PU21 encoding function, several legacy video quality models such as SSIM \cite{ref:ssim}, MS-SSIM \cite{ref:ms_ssim}, FSIM \cite{ref:fsim}, and VSI \cite{ref:vsi} have been shown to be useful for HDR video quality assessment problems \cite{ref:live_hdr}.

A recent method of adapting SDR quality models for HDR quality prediction is the non-linear HDRMAX transformation. Two variants of HDRMAX have been proposed in the literature, which we shall refer to as HDRMAX1 \cite{ref:hdrmax} and HDRMAX2 \cite{ref:live_hdr}, which respectively deploy a single and a double non-linearity. Both variants of HDRMAX operate as a side channel of a standard video quality model, by extracting quality-aware features on emphasized dark and bright video frame regions. These regions are the primary contributors to improved HDR video quality relative to SDR quality, but these regions are also susceptible to highly visible distortions that are poorly captured by SDR video quality models. HDRMAX2 has been shown to improve the accuracy of VMAF \cite{ref:live_hdr}, while HDRMAX1 has been shown to significantly improve the prediction accuracies of a wide range of FR and no-reference (NR) video quality models \cite{ref:hdrmax}.

In this way, modern HDR video quality predictors derive their success, in part, from successful SDR models. Within the space of quality models designed to predict the perceptual qualities of SDR videos that have been subjected to compression, the FUNQUE+ \cite{ref:funque_plus} suite of models has been shown to achieve SOTA accuracy. FUNQUE+ models are based on the FUNQUE \cite{ref:funque} framework, which enables the development of efficient and accurate video quality models. Efficient and accurate designs are achieved by the use of a shared perceptually-sensitive wavelet transform space within which all quality features are computed.

The remainder of the paper is organized as follows. Section \ref{sec:background} reviews the FUNQUE+ suite of models and the HDRMAX transformation, which represent SOTA performance among SDR and HDR video quality assessment models. The experimental methodology used to evaluate FUNQUE+ for HDR is reviewed in Section \ref{sec:experiments}, and the results of the evaluation are presented in Section \ref{sec:results}. Finally, we provide concluding remarks in Section \ref{sec:conclusion}.
\section{Bakground}
\label{sec:background}

\subsection{FUNQUE+}
Our approach to the design of FR video quality prediction models relies on the FUNQUE \cite{ref:funque, ref:funque_plus} framework, which yields compact and efficient quality models. The defining feature of FUNQUE is the use of a ``unified'' transform, which is a perceptually sensitized wavelet transform that is shared by a set of ``atom quality models`` that provide quality-aware features. The computation sharing provided by the unified transform greatly improves model efficiency, while the introduction of perceptual sensitivity improves model accuracy.

A key component of the unified transform is the use of the Self-Adaptive Scale Transform (SAST) \cite{ref:sast}, whereby video frames to be quality-analyzed are first scaled to account for the viewing distance relative to the display size. The SAST scale factor can be approximated \cite{ref:funque_plus} as
\begin{equation}
    \alpha_{SAST} \approx \frac{D/H}{1.618},
\end{equation}
where \(D/H\) is the ratio of the viewing distance \(D\) to the display height \(H\). By assuming that 1080p displays are viewed at a distance of \(3H\) \cite{ref:viewing}, the value \(\alpha_{SAST} \approx 2\) was obtained and used in \cite{ref:funque} and \cite{ref:funque_plus}. However, 4K displays are often assumed to be placed at a viewing distance 1.5 times the display height \cite{ref:viewing}. This ratio was loosely enforced in the subjective experiments used to build the LIVE-HDR database \cite{ref:live_hdr}. Hence, for 4K viewing, \(\alpha_{SAST} \approx 1\) and SAST is implicitly included in FUNQUE+ models when applied at 4K resolution.

The FUNQUE framework was refined in \cite{ref:funque_plus}. resulting in the FUNQUE+ suite of models, by the use of a variety of Contrast Sensitivity Functions (CSFs), and by adapting existing quality models to be computed in the unified transform space, including VIF \cite{ref:vif}, ST-RRED \cite{ref:strred}, and the Spatial Activity Index (SAI) \cite{ref:tlvqm}. A novel Multi-Scale Enhanced SSIM (MS-ESSIM) \cite{ref:essim} feature forms the backbone of FUNQUE+ models. \cite{ref:funque_plus}. In a substantial cross-database evaluation study, the Y-FUNQUE+ and 3C-FUNQUE+ models developed in \cite{ref:funque_plus} were shown to be the best luma-only and three-channel (i.e., including chroma) FR models. 

Here, we evaluate the efficacy of these two models when applied to the task of HDR video quality prediction, using the LIVE-HDR \cite{ref:live_hdr} database. The feature sets of the two models are presented in Table \ref{tab:funque_plus_feats}. The names of the features in the descriptions of the FUNQUE+ models follow the notation in \cite{ref:funque_plus}, and the prefix ``Y-'', ``Cb-'', or ``Cr-'' denotes the channel from which the corresponding feature is computed.

A short description of the features referenced in Table \ref{tab:funque_plus_feats} is as follows. We refer the reader to \cite{ref:funque_plus} for a more detailed exposition of the FUNQUE+ feature set.

\begin{itemize}
    \item MS-ESSIM: A multi-scale version of the Enhanced SSIM (ESSIM) algorithm developed in \cite{ref:essim}.
     ESSIM differs from SSIM through the use of SAST, the use of small square windows for local moment computation, and the use of Coefficient of Variation (CoV) pooling instead of the traditional average pooling. Inspired by ESSIM, SAST was incorporated into all features used by the FUNQUE+ models. ESSIM was adapted to MS-ESSIM using an exponentially weighted product similar to the design of MS-SSIM \cite{ref:ms_ssim}.
     \item MAD-Ref: The mean absolute difference (MAD) between approximation subbands of the perceptually sensitized wavelet decompositions of successive frames from the reference video. MAD-Ref is a measure of the temporal complexity of the source video.
     \item DLM-S: A single-scale version of the detail loss metric (DLM) \cite{ref:dlm} computed at the coarsest scale of the perceptually sensitized wavelet decomposition of the reference and test videos.
     \item MAD-Dis: Similar to MAD-Ref, but computed from the distorted, i.e., test video.
     \item SRRED, TRRED: Spatial and Temporal components of a scalar version of the spatiotemporal reduced-reference entropic difference (ST-RRED) model \cite{ref:strred}. ST-RRED is a spatiotemporal quality model that utilizes local entropies of wavelet coefficients to predict perceptual quality.
     \item Edge: The ``Edge'' feature proposed in \cite{ref:enh_vmaf} measures the degree of edge enhancement in the test video, relative to the reference video, by analyzing the absolute differences between wavelet coefficients.
     \item MAD: Similar to MAD-Ref and MAD-Dis, but computed as the difference between approximation subbands of the wavelet decompositions of corresponding reference and test video frames.
\end{itemize}

\begin{table*}[b]
    \centering
    \caption{Feature Sets of FUNQUE+ Models}
    \label{tab:funque_plus_feats}
    \begin{tabular}{|c|c|}
    \hline
    Model & Features \\
    \hline
    Y-FUNQUE+ & \(\text{Y-MS-ESSIM} + \text{Y-MAD-Ref} + \text{Y-DLM-S}\) \\
    \hline
    3C-FUNQUE+ & \(\text{Y-MS-ESSIM} + \text{Y-MAD-Dis} + \text{Y-DLM-S} + \text{Y-SRRED-HV} + \text{Y-TRRED-HV} + \text{Cb-Edge} + \text{Cr-MAD} \) \\
    \hline
    \end{tabular}
\end{table*}
\subsection{HDRMAX}
\label{sec:hdrmax}
HDRMAX is a suite of non-linear preprocessing methods designed to adapt SDR video quality prediction models to HDR. Two variants of HDRMAX have been proposed, which we refer to as HDRMAX1 \cite{ref:hdrmax} and HDRMAX2 \cite{ref:live_hdr}, corresponding to the use of one or two non-linearities. Both methods apply a local normalization operation followed by one or more non-linear transforms. The goal of both HDRMAX variants is to emphasize the measurement of distortion on the bright and dark regions of HDR video frames, enabling video quality models to make much more accurate predictions of HDR video quality. Indeed, in \cite{ref:hdrmax}, HDRMAX has been reported to improve the median SRCCs of FR VQA algorithms by 25\% on the LIVE HDR database.

HDRMAX1 normalizes the local luminance of images \(I(i,j)\) to the range \([-1, 1]\) using min-max normalization. In \cite{ref:hdrmax}, normalized luminance is computed within 17\(\times\)17 neighborhoods via
\begin{equation}
    \Tilde{I}_{minmax}(i,j) = 2\left(\frac{I(i,j) - I_{min}(i,j)}{I_{max}(i,j) - I_{min}(i,j)}\right) - 1,
\end{equation}
then subjected to the double exponential non-linearity
\begin{equation}
    \mathrm{HDRMAX1}(x) = \mathrm{sgn}(x) \exp\left(4\mid x\mid\right) - 1.
\end{equation}

By contrast, HDRMAX2 uses local mean subtraction to normalize the local luminance of images. The locally weighted mean is computed within 31\(\times\)31 Gaussian-weighted windows and subtracted from the luminance values:

\begin{equation}
    \Tilde{I}_{meansub}(i,j) = I(i,j) - I_{mean}(i,j) \quad \forall (i,j)
\end{equation}

The mean-normalized luminance values are then transformed using two exponential non-linearities, one that emphasizes bright regions, while the other emphasizes dark regions:

\begin{equation}
    \mathrm{HDRMAX2}_{pos}(x) = \exp\left(0.5x\right)
\end{equation}
\begin{equation}
    \mathrm{HDRMAX2}_{neg}(x) = \exp\left(-5x\right)
\end{equation}

\begin{figure*}[t]
    \centering
    \subfloat[0.33\linewidth][\(\mathrm{HDRMAX1}\)]{\includegraphics[width=0.33\linewidth]{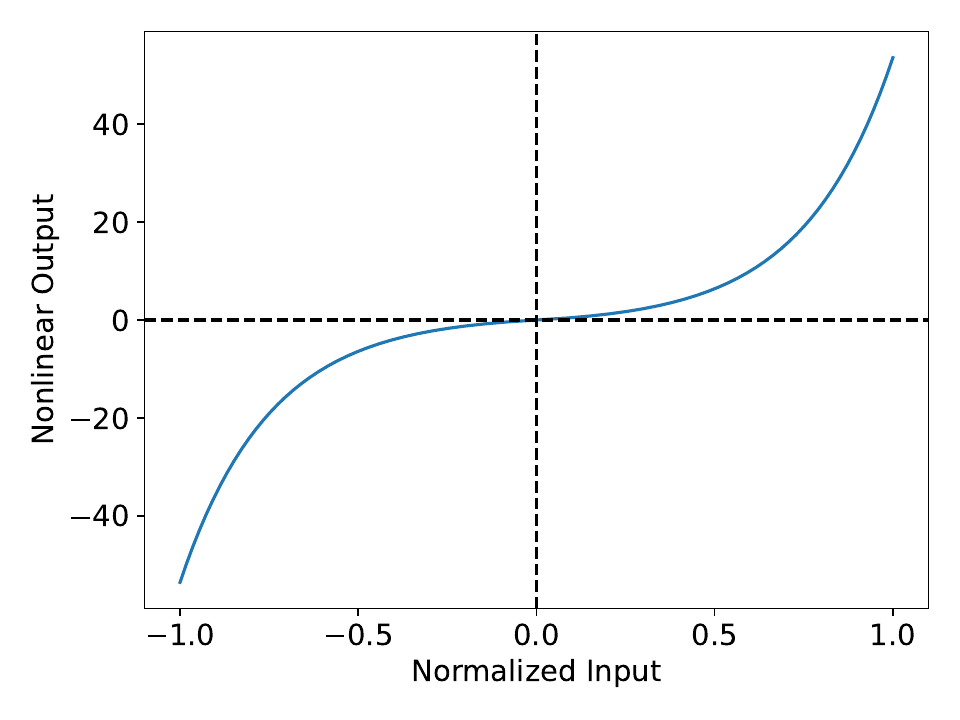}}%
    \subfloat[0.33\linewidth][\(\mathrm{HDRMAX2}_{pos}\)]{\includegraphics[width=0.33\linewidth]{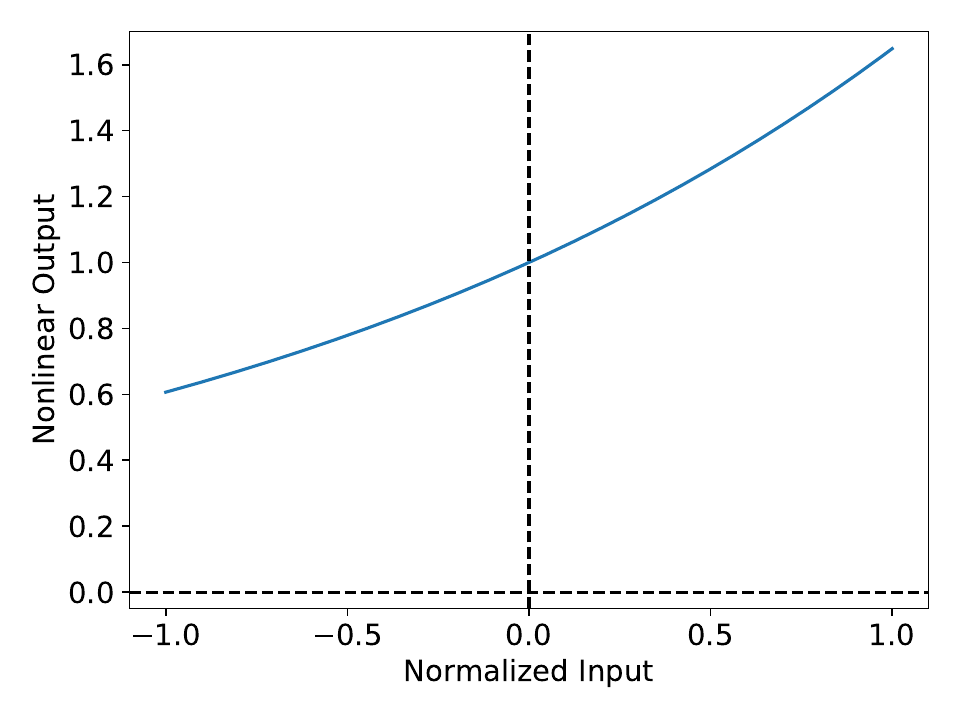}}%
    \subfloat[0.33\linewidth][\(\mathrm{HDRMAX2}_{neg}\)]{\includegraphics[width=0.33\linewidth]{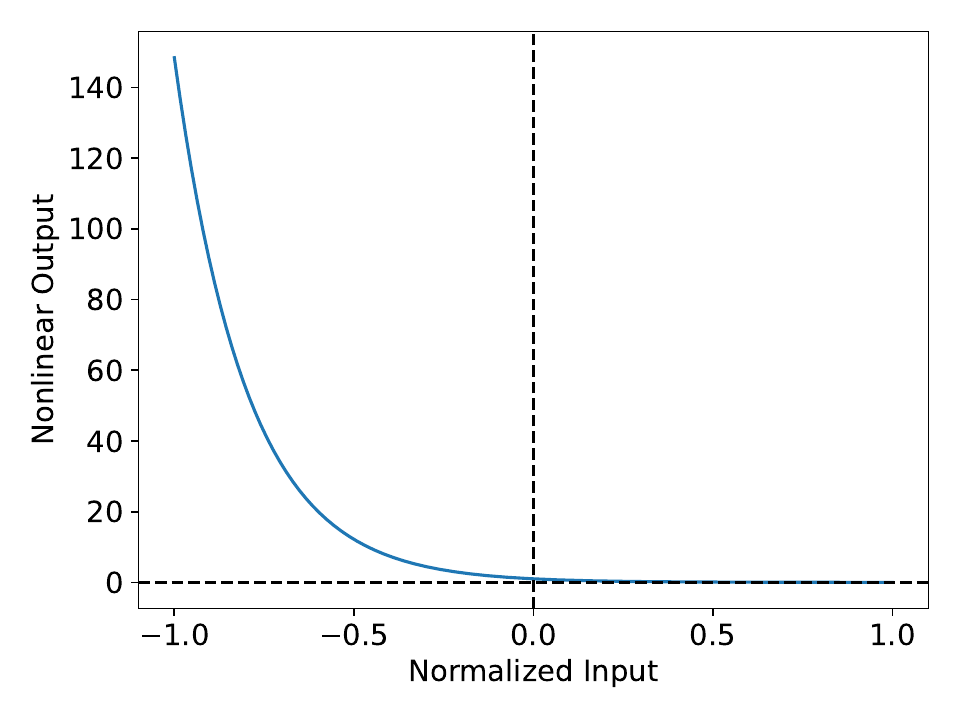}}%
    \caption{The HDRMAX Non-linearities}
    \label{fig:hdrmax}
\end{figure*}

The three HDRMAX non-linearities are illustrated in Fig.~\ref{fig:hdrmax}.

\section{Experiments}
\label{sec:experiments}

\subsection{The LIVE-HDR Database}
The LIVE-HDR database \cite{ref:live_hdr} is a subjectively annotated database of 310 compressed 4K HDR videos. The distorted videos in the database were obtained by encoding 31 unique 10-bit HDR source contents using HEVC at ten different compression levels. All of the source contents were generated at professional studios and conform to the PQ HDR10 standard.

The subjective annotations were obtained by presenting the videos to human subjects under two ambient brightness conditions - bright and dark. Although it was found that the two ambient conditions yielded statistically similar subjective quality scores, we evaluated all of the compared models on both sets of subjective scores, since the accuracies of video quality models have been shown to vary between the two scenarios \cite{ref:live_hdr}.

\subsection{Evaluation}
We utilized content-separated random cross-validation to evaluate the compared quality models and to tune regressor hyper-parameters. Specifically, 1000 random splits of the dataset were generated ensuring that the same source content was not present in both the training and test splits. The Pearson Correlation Coefficient (PCC), Spearman Rank Order Correlation Coefficient (SROCC), and Root Mean Squared Error (RMSE) were computed between model predictions and subjective ratings under the bright and dark ambient conditions to measure model accuracy, and the median value over random splits was reported as the overall accuracy. Regressor hyper-parameters were tuned to optimize the average of PCC and SROCC for both bright and dark ambient conditions. Following the recommendations in \cite{ref:live_hdr}, we ensured that there was no ``leakage'' of football and golf videos from multiple source clips between the train and test splits.

In addition to the ``base'' models, we also evaluated the effect of HDRMAX on the two FUNQUE+ models. In \cite{ref:live_hdr} and \cite{ref:hdrmax}, the ``HDRMAX feature set'' involves computing the VMAF feature set (excluding motion) from HDRMAX-transformed input frames. We followed this theme by computing HDRMAX features from input frames transformed using both HDRMAX1 and HDRMAX2. In this protocol, the size of the feature set was always increased by 5 by HDRMAX1 and 10 by HDRMAX2. Finally, we compared the accuracies of the FUNQUE+ models against SOTA FR quality models on the LIVE-HDR database.
\section{Results}
\label{sec:results}
In this section, we present the results of evaluating various quality models on the LIVE-HDR database. These models include SDR quality models such as PSNR, SSIM \cite{ref:ssim}, MS-SSIM \cite{ref:ms_ssim}, etc., that have been adapted to HDR using PU21 \cite{ref:pu21}, the visible difference predictor (VDP) models HDR-VDP versions 2.2 \cite{ref:hdr_vdp_2} and 3.0.7 \cite{ref:hdr_vdp_3}, a recent deep learning model named HIDRO-FR \cite{ref:hidro}, and the fusion based models VMAF \cite{ref:vmaf}, Y-FUNQUE+ \cite{ref:funque_plus}, and 3C-FUNQUE+ \cite{ref:funque_plus}. For the three fusion models, we also consider feature sets that have been augmented using both HDRMAX1 and HDRMAX2.

The median PCC, SROCC, and RMSE achieved over 1000 random train-test splits are reported in Table \ref{tab:sota_comp}. The three best results in each column are boldfaced. It may be seen that the 3C-FUNQUE+ models, including those augmented by HDRMAX1 and HDRMAX2, achieve SOTA accuracy. In particular, these models achieve the best generalization between both dark and bright ambient conditions among all compared models.

Moreover, we observe that even without HDRMAX augmentation, both the Y-FUNQUE+ and 3C-FUNQUE+ feature sets still achieve similar or higher accuracies than the best version of VMAF, which is augmented with HDRMAX2. This is particularly notable since VMAF+HDRMAX2 contains a total of 16 features, while Y-FUNQUE+ and 3C-FUNQUE+ contain three and seven features respectively. This demonstrates the superiority of their ``base'' FUNQUE+ feature sets. 

\begin{table*}[t]
    \centering
    \caption{Comparing HDR FUNQUE+ Models Against SOTA FR Quality Models}
    \label{tab:sota_comp}
    \begin{tabular}{|c|c|c|c|c|c|c|}
        \hline
        \multirow{2}{*}{Model} & \multicolumn{3}{c|}{Dark Ambient} & \multicolumn{3}{c|}{Bright Ambient} \\
        \cline{2-7}
         & SROCC \(\uparrow\) & PCC \(\uparrow\) & RMSE \(\downarrow\) & SROCC \(\uparrow\) & PCC \(\uparrow\) & RMSE \(\downarrow\) \\
        \hline
        PU21-PSNR \cite{ref:pu21} & 0.5841 & 0.5767 & 14.2798 & 0.6117 & 0.5963 & 13.9762 \\
        HDR-VDP-2.2 \cite{ref:hdr_vdp_2} & 0.5868 & 0.5128 & 15.0052 & 0.6472 & 0.6254 & 13.986 \\
        PU21-SSIM \cite{ref:pu21, ref:ssim} & 0.6019 & 0.6065 & 13.8971 & 0.6403 & 0.6301 & 13.5188 \\
        PU21-FSIM \cite{ref:pu21, ref:fsim} & 0.6470 & 0.6372 & 13.4705 & 0.7116 & 0.6904 & 12.5951 \\
        PU21-MSSSIM \cite{ref:pu21, ref:ms_ssim} & 0.6593 & 0.6564 & 13.1868 & 0.7120 & 0.6969 & 12.4859 \\
        PU21-VSI \cite{ref:pu21, ref:vsi} & 0.6795 & 0.6667 & 13.0284 & 0.7290 & 0.7058 & 12.3334 \\
        HDR-VDP-3.0.7 \cite{ref:hdr_vdp_3} & 0.7363 & 0.7307 & 11.9332 & 0.8080 & 0.8098 & \textbf{10.2139} \\
        HIDRO-FR \cite{ref:hidro} & 0.8673 & 0.8400 & \textbf{9.5010} & \textbf{0.8929} & \textbf{0.8640} & \textbf{9.1699} \\
        VMAF \cite{ref:vmaf} & 0.8123 & 0.7352 & 17.7120 & 0.8572 & 0.7853 & 17.0148 \\
        VMAF + HDRMAX-1 \cite{ref:vmaf, ref:hdrmax} & 0.8109 & 0.7461 & 13.9312 & 0.8497 & 0.7809 & 13.1947 \\
        VMAF + HDRMAX-2 \cite{ref:vmaf, ref:live_hdr} & 0.8530 & 0.8038 & 13.3552 & 0.8877 & 0.8264 & 12.7438 \\
        Y-FUNQUE+ \cite{ref:funque_plus} & 0.8720 & 0.8301 & 11.2602 & 0.8709 & 0.8218 & 11.6188 \\
        Y-FUNQUE+ + HDRMAX-1 \cite{ref:funque_plus, ref:hdrmax} & 0.8739 & 0.8326 & 10.9248 & 0.8772 & 0.8276 & 11.8800 \\
        Y-FUNQUE+ + HDRMAX-2 \cite{ref:funque_plus, ref:live_hdr} & 0.8579 & 0.8080 & 12.4202 & 0.8722 & 0.8296 & 15.9837 \\
        3C-FUNQUE+ \cite{ref:funque_plus} & \textbf{0.8985} & \textbf{0.8732} & \textbf{9.4463} & 0.8895 & \textbf{0.8576} & \textbf{9.8930} \\
        3C-FUNQUE+ + HDRMAX-1 \cite{ref:funque_plus, ref:hdrmax} & \textbf{0.9004} & \textbf{0.8735} & 10.5711 & \textbf{0.8896} & 0.8524 & 10.9588 \\
        3C-FUNQUE+ + HDRMAX-2 \cite{ref:funque_plus, ref:live_hdr} & \textbf{0.9022} & \textbf{0.8738} & \textbf{9.2223} & \textbf{0.8906} & \textbf{0.8583} & 10.5827 \\
        \hline
    \end{tabular}
\end{table*}

\section{Conclusion}
\label{sec:conclusion}
We have reviewed the FUNQUE+ suite of models that were designed for the quality assessment of compressed SDR videos, and successfully demonstrated their applicability to HDR VQA. Specifically, we evaluated the accuracy of the Y-FUNQUE+ and 3C-FUNQUE+ fusion-based FR VQA models, both with and without recently proposed HDRMAX augmentations. We found through extensive cross-validation studies that the FUNQUE+ models matched or surpassed SOTA full-reference quality models including deep methods.

A key limitation of this work, and indeed similar work on compressed HDR VQA, is the availability of only a single dataset for model evaluation. As a result, we are unable to evaluate the cross-database generalization capacity of any of the models under consideration. The development of more subjective databases is critical to overcoming this limitation. Such additional data could be used to tune FR quality feature sets to better reflect HDR video properties, such as the importance of chroma distortions and distortions across luminance ranges.

\section{Acknowledgment}
\label{sec:acknowlegment}
The authors acknowledge the Texas Advanced Computing Center (TACC) at The University of Texas at Austin for providing high-performance computing (HPC) resources that have contributed to the research results reported within this paper. URL: \url{http://www.tacc.utexas.edu}. The authors would also like to thank the authors of \cite{ref:hidro} for providing supplementary results of their model for use here.

\bibliographystyle{IEEEbib}
\bibliography{refs}

\end{document}